# GUARD-D-LLM: An LLM-Based Risk Assessment Engine for the Downstream uses of LLMs

Sundaraparipurnan Narayanan

Doctoral Candidate, University of Bordeaux, sundar.narayanan@aitechethics.com

Sandeep Vishwakarma,

Masters in Artificial Intelligence, Symbiosis Institute of Technology, sandeep.vishwakarma.mtech2022@sitpune.edu.in

Amidst escalating concerns about the detriments inflicted by AI systems, risk management assumes paramount importance, notably for high-risk applications as demanded by the European Union AI Act. Guidelines provided by ISO and NIST aim to govern AI risk management; however, practical implementations remain scarce in scholarly works. Addressing this void, our research explores risks emanating from downstream uses of large language models (LLMs), synthesizing a taxonomy grounded in earlier research. Building upon this foundation, we introduce a novel LLM-based risk assessment engine (GUARD-D-LLM: Guided Understanding and Assessment for Risk Detection for Downstream use of LLMs) designed to pinpoint and rank threats relevant to specific use cases derived from text-based user inputs. Integrating thirty intelligent agents, this innovative approach identifies bespoke risks, gauges their severity, offers targeted suggestions for mitigation, and facilitates risk-aware development. The paper also documents the limitations of such an approach along with way forward suggestions to augment experts in such risk assessment thereby leveraging GUARD-D-LLM in identifying risks early on and enabling early mitigations. This paper and its associated code serve as a valuable resource for developers seeking to mitigate risks associated with LLM-based applications.

## 1 INTRODUCTION

Risk management is a tool to systematically (1) promote responsible business practices, (2) align with the focus on ethical considerations, (3) function as an essential pillar of the auditing process, and (4) support the existing industry specific measures of oversight. In the context of artificial intelligence (AI) and machine learning (ML), risk management is essential for ensuring responsible and trustworthy AI deployment. Organizations like ISO and NIST have developed frameworks such as ISO 31000 and NIST's AI risk management framework, which are crucial for addressing AI and ML risks [2] [27] [21] [35] [41] [45]. EU AI Act has also required adoption of tailored risk management mechanism for high-risk AI systems. These frameworks (ISO/ NIST) align with the focus on ethical considerations, bias mitigation, and safety issues in AI and ML [39] [56]. In addition, Risk management plays an integral role in facilitating the process of auditing AI and ML systems, helping identify vulnerabilities and ethical concerns [16]. It also support specific domains, such as managing AI risks in financial services [14], managing AI risks in healthcare [28], high-consequence AI risk management [4], supply chain risk management [5], in identifying AI-specific risks [44] [3], and AI governance [16]. It therefore plays a pivotal role in ensuring the reliability, safety, and ethical use of AI systems across various applications.

## 1.1 Emergence of LLMs

The emergence of Large Language Models (LLMs) has revolutionized the field of natural language processing, leading to significant advancements in various downstream tasks. Notable LLMs such as GPT4 [30], LLama2 [46], Falcon [34], and Mistral [17] have garnered attention due to their adaptability and effectiveness across diverse applications. Research focusing on open-sourced LLMs [7] has contributed to democratizing advanced language models, enabling widespread access and utilization.

These LLMs have been instrumental in driving innovations in several key areas, including adapting LLMs, prompt engineering, emergent abilities, and optimization strategies. In the area of adaption of LLMs, advanced techniques such as leveraging large context length [9] [18] [37], and fine-tuning methods like instruction tuning [54], alignment tuning, & memory-efficient model adaptation [1] [20] [25] have emerged. These approaches allow LLMs to be fine-tuned for specific tasks and domains, enhancing their performance and applicability to targeted applications.

Additionally, prompt engineering techniques such as in-context learning [22], chain-of-thought [47] [48], tree-of-thought [52], and planning have facilitated the development of more sophisticated and context-aware language models [23] [51] [55]. These techniques enable LLMs to understand and generate language in a more nuanced and coherent manner, enhancing their practical utility through efficient adaptation to various contexts.

Moreover, emergent abilities such as in-context learning, instruction following, and step-by-step reasoning (CoT) have expanded the capabilities of LLMs, enabling them to comprehend and reason through complex language inputs effectively. Fine-tuning methods [9] [31] [36] [53] play a crucial role in tailoring these emergent abilities to specific tasks, ensuring optimal performance in diverse scenarios.

Furthermore, optimization strategies, including prompt optimization techniques like vLLM [20] or LLMLingua , and inference optimization strategies, have enhanced the efficiency and performance of LLMs in processing and generating language. Efficient approaches to fine-tuning contribute significantly to the optimization of these models, ensuring that they can adapt quickly and effectively to changing requirements.

## 1.2 Need for downstream LLM risk management

Multiple studies have explored the potential risks and applications of large language models (LLMs). The utilization of Large Language Models (LLMs) in downstream applications carries various risks, including privacy concerns where these models can capture sensitive information in text embeddings. [49] [50] and [29] have identified various ethical and social risks, including discrimination, misinformation, and misuse. [32] has specifically highlighted the risk of information pollution through misinformation.

# 2 ESTABLISHING BASE TAXONOMY FOR LLM RISK ASSESSMENT

## 2.1 Existing risk and harm taxonomies

Several papers propose various taxonomies and frameworks to address security, ethical, and social risks associated with Large Language Models (LLMs) and enhance their safety and trustworthiness. [50] and [49] discussed ethical and social risks. [12] classified attacks on LLMs, highlighting the need for improved security solutions. [26] focused on security threats, including fraud and cybercrime, while Europol emphasizes the need for safeguards against misuse. [43] provided a taxonomy for sociotechnical harms in algorithmic systems but may face challenges in anticipating all possible consequences. Further, [24] covered various dimensions of trustworthiness in LLMs. [38] proposed benchmarks to evaluate harmful text generation by LLMs. [10] discussed the need for comprehensive taxonomies and benchmarks for LLM systems.

While none of the taxonomies were developed with a focus on downstream uses of LLMs, the taxonomies apply to many circumstances. However, there is a need to have a specific approach to explore the risks associated with downstream uses of LLM, which may be different and contextual to the use case themselves.

## 2.2 Approach towards establishing base taxonomy

To that end, a three-dimensional approach and a broad listing of relevant risks and their sub-risks are provided in this paper. The approach is not called a taxonomy as this taxonomy does not offer anything new, but just represents the risks in an alternative way. The three-dimension approach focuses on (a) Process Risk, (b) Component Risk, and (c) Use Case Risk. The process risk covers the risks associated with data aggregation, embedding, prompt engineering, Retrieval augmented generation, Fine-tuning, evaluation, validation, moderation, and monitoring. The component risk covers data, model, pipeline, infrastructure, interface, integrations, deployment, and human-in-the-loop. The use case risk covers risks arising from the scope, nature, context, and purpose of the use case itself. Overall, there are 30 broad risks and 95 sub risks that are compiled as part of the exercise. By enabling analysis of the risks through these three lenses, and by providing an illustrative list of risk categories, the framework offers a comprehensive methodology for understanding foreseeable, emergent, and systemic societal risks and supports in adopting informed risk mitigation strategies across the entire downstream LLM ecosystem.

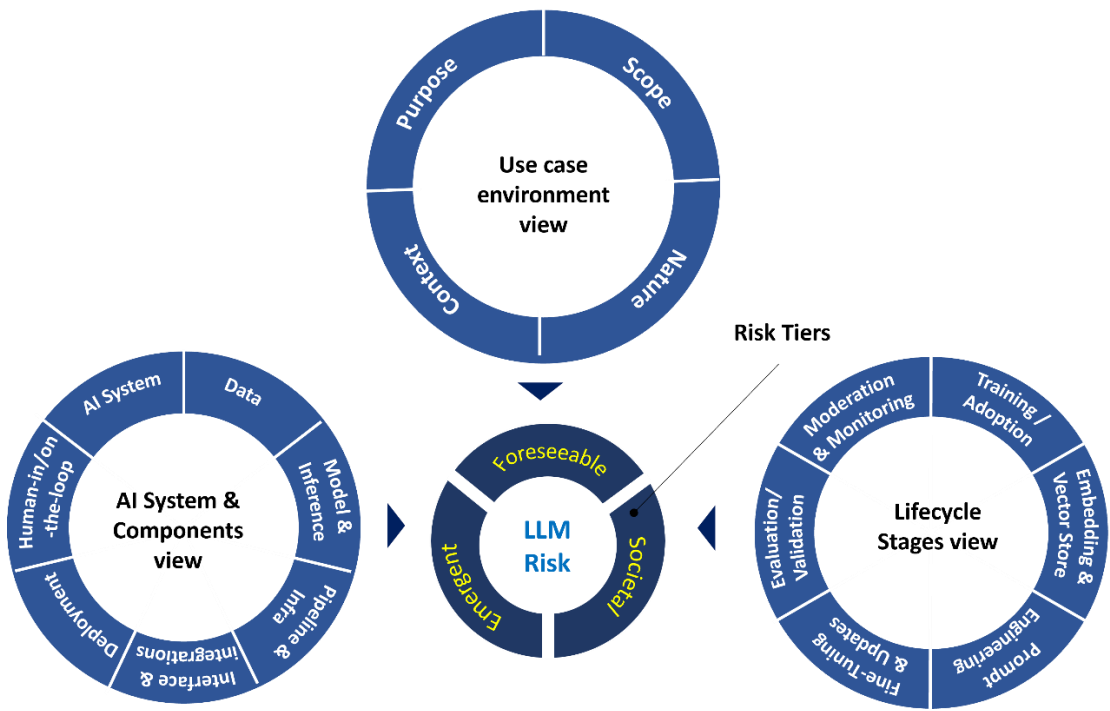

Figure 1: An overall approach representing the three dimensional view of LLM downstream risk

## 2.3 Risk landscape from the lens of downstream use cases

The broad risks compiled are classified under 4 categories, namely, (A) Content Risks, (B) Context Risks, (C) Trust Risks and (D) Societal and Sustainability Risks. The details of the same are explained below:

**Broad Risks and Sub Risks**

| 1. Content Risks | 2. Context Risks | 3. Trust Risks | 4. Societal Impact and Sustainability Risks |
|---|---|---|---|
| 1.1 Toxic or harmful content<br>1.2 Incorrect or inaccurate content<br>1.3 Propagating misconceptions/false beliefs/ Unfaithful content<br>1.4 Dissemination of dangerous information<br>1.5 Fraudulent suggestions or information<br>1.6 Manipulative or persuasive content | 2.1 Unethical use<br>2.2 Unfair performance/capability distribution<br>2.3 Influence operations to manipulate users/ people<br>2.4 Overreliance or automation bias<br>2.5 Exploitative data sourcing and enrichment<br>2.6 False representation of performance<br>2.7 Lack of disclosure of automation/ LLMs use | 3.1 Lack of accountability<br>3.2 Inadequate explainability<br>3.3 Violation of personal integrity<br>3.4 Misappropriation or exploitation or data/ information<br>3.5 Exposure to intellectual property<br>3.6 Safety exposure<br>3.7 Security threats<br>3.8 Privacy infringement<br>3.9 Insufficient safeguards | 4.1 Environmental damage<br>4.2 Inequality or precarity<br>4.3 Undermine creative economies<br>4.4 Unfair representation or Stereotypes<br>4.5 Discrimination or bias<br>4.6 Defamation<br>4.7 Pollution of the information ecosystem<br>4.8 Unfair distribution of benefits/ information |

Figure 2: Risks and sub-risks associated with the four broad categories are listed above

### 2.3.1 Content Risks

Content risks encompass a wide range of challenges associated with the generation and dissemination of digital content, especially through language models. These risks include the creation and promotion of toxic or harmful content, such as hate speech, radical ideologies, and cyberbullying, which can negatively impact individuals and communities. Additionally, content risks extend to the production of incorrect or inaccurate information, including misinformation and misleading answers, which can contribute to the spread of false beliefs and misconceptions. Moreover, content risks encompass the dissemination of dangerous information, such as terrorist propaganda and fraudulent suggestions, which pose threats to public safety and user well-being. Finally, manipulative or persuasive content risks involve the unethical use of language models to influence emotions, beliefs, and behaviors, including political manipulation and encouraging unethical actions, raising concerns about ethical and societal implications.

### 2.3.2 Context Risks

Context risks refer to the potential negative consequences arising from the application of large language models (LLMs) in specific situations or contexts. These risks encompass unethical use, such as generating deceptive or fraudulent content and manipulating public opinion through LLM-generated messages. Additionally, context risks include issues like unfair distribution of LLM capabilities, influence operations for malicious purposes, overreliance on automated decision-making without human oversight, exploitative data sourcing practices, false representation of LLM capabilities, failure to address limitations, and the lack of transparency in disclosing the use of LLMs in various interactions. These risks highlight the importance of ethical and responsible deployment of LLMs to prevent harm, bias, and misuse in diverse contexts.

*2.3.3 Trust Risks*

Trust risks encompass a wide range of concerns, including accountability, responsibility, legality, and transparency. There's a lack of clear mechanisms to hold developers, users, or platform operators accountable for language model outputs, leading to ambiguous responsibility and legal gaps in addressing potential harms. Additionally, issues related to explainability and transparency further erode trust, as language models often produce black-box outputs and unexplained decisions, which can have serious implications for personal integrity and privacy. Inadequate safeguards, such as content filtering and monitoring, exacerbate these trust risks, emphasizing the need for a more comprehensive and accountable approach to the development and deployment of language models.

*2.3.4 Societal and Sustainability Risks*

Societal Impact and Sustainability Risks in the context of large language models involve several critical concerns. These risks encompass environmental damage resulting from the energy-intensive training and deployment of such models in data centers, leading to an increased carbon footprint. Additionally, there's the risk of exacerbating inequality and precarity through biased outputs and the concentration of benefits, which can further reinforce social disparities. Furthermore, the potential undermining of creative economies due to automated content generation threatens the livelihoods of professionals in various industries. The amplification of unfair representations and stereotypes, as well as the risk of discrimination and defamation, further highlight the profound societal impacts of large language models, emphasizing the need for responsible development and usage to mitigate these risks and foster sustainability and equity.

## 3 USING LLM FOR RISK MANAGEMENT

### 3.1 Existing uses of LLMs in risk handling or associated fields

Use of LLM for risk management has not been widely explored in research. However, the application of Large Language Models (LLMs) in the domain of software testing has garnered substantial attention within the computer science community in recent years [15]. Various research have delved into distinct aspects of this intersection, demonstrating the versatility of LLMs in enhancing different facets of software testing methodologies. Recent studies have explored novel applications of LLMs in software testing. These include investigations into zero-shot vulnerability repair using LLMs [34], the development of adaptive test generation techniques leveraging LLMs [40], and the automation of unit testing through the incorporation of LLMs. The detection of harmful language has been explored through the application of LLMs with counterfactually augmented data [42], and LLMs have been employed in automatic penetration testing [11]. Furthermore, [13] has explored the use of LLMs in hazard analysis and [19] has provided a survey of methods for addressing potential threats and societal harms from language generation models, while [6] explored use of LLMs in assessing risk of suicide.

### 3.2 Methodology for developing LLM as a risk assessment engine

The new template enables you to import required indexing concepts for your article from
To explore using LLM inference to conduct risk assessment, a risk assessment engine was developed titled GUARD-D-LLM. In the development of the GUARD-D-LLM (Guided Understanding and Assessment for Risk Detection for Downstream use of LLMs) tool, we have outlined a systematic methodology for risk assessment tailored to specific use cases derived from textual user inputs. This methodology consists of the following steps:

*3.2.1 Information gathering*

The first step involves collecting comprehensive information related to the downstream use of LLMs. This encompasses various aspects such as data sources, model specifications, user demographics, use case objectives, LLM characteristics, embedding methods, prompt engineering, fine-tuning techniques, monitoring and moderation strategies, deployment models, and feedback mechanisms. Users provide the use case title and brief information, triggering the appearance of a dynamic table that specifies additional information required for the assessment. Updating these responses initiates the risk assessment process. The perspective on risk when implementing downstream applications with Large Language Models (LLMs) underscores the multifaceted nature of these considerations. The versatility of LLMs in various functions brings both opportunities and risks, emphasizing the need to assess and manage potential drawbacks carefully. The diversity of data types and user groups necessitates a comprehensive understanding of data sourcing, handling, and user requirements. Furthermore, the purpose of the use case plays a crucial role in risk assessment, as different goals may entail varying degrees of risk. The type of LLM used and the embedding techniques employed influence the behavior and performance of the AI system, adding complexity to risk evaluation. Guardrails and evaluation methods are essential for responsible development, but monitoring and moderation are equally critical to ensure ongoing responsible operation.

*3.2.2 Risk assessment by agents*

The risk assessment is conducted using LLMs, focusing on the ethical, regulatory, and governance challenges associated with LLM deployment in line with Responsible AI principles. Thirty context-aware risk agents are deployed based on the compiled risks from the previous step.

*3.2.3 Dynamic risk collection agent*

An agent is designed to identify dynamic risks by conducting web searches related to user-provided considerations such as limitations, drawbacks, disadvantages, and failures of tools, libraries, datasets, or utilities and compile top three issues for each consideration. This agent generates the risks and assesses their contextual relevance to the use case. Similar to the risk assessment agents, this agent also evaluates risks for severity, likelihood, and preliminary risk scores at an agent level.

*3.2.4 Risk compilation and refinement*

The risks identified by the risk assessment agents and the dynamic risk collection agent are compiled. Low or negligible risk cases are eliminated, and the remaining risks are reranked to create a use case-specific risk ranking at an overall level.

*3.2.5 Mitigation measures*

For each identified risk, the GUARD-D-LLM suggests governance measures aimed at providing guidance to users on how to mitigate these risks effectively.

*3.2.6 Report generation*

The information gathered and analyzed throughout the preceding steps is compiled into a comprehensive report that users can download. This report serves as a valuable resource, offering insights into the identified risks, their severity, and recommended governance measures, facilitating informed decision-making and risk management in the downstream use of LLMs.

The above approach was adopted and the code relating to the same was pushed to Github. Preliminary assessments using GUARD-D-LLM revealed certain limitations and challenges as referred in the next section.

## 4 DISCUSSION

### 4.1 Limitations of the current approach

In the context of Large Language Model (LLM)-based risk assessment for downstream applications, several critical limitations must be acknowledged. Firstly, the effectiveness of such assessments heavily hinges on the availability of comprehensive and accurate information pertaining to the specific use case. Incomplete or biased information may result in inaccurate risk evaluations. Furthermore, LLMs may encounter challenges when tasked with assessing risks in intricate, highly complex, or niche use cases that demand specialized domain knowledge or intricate contextual factors.

Another significant limitation lies in the dynamic nature of risks associated with LLM applications. These risks evolve over time, and LLM-based assessments may not promptly capture emerging threats or vulnerabilities. Additionally, LLMs lack domain-specific expertise and may fail to recognize nuanced risks that necessitate specialized knowledge within a particular field. They may also struggle to accurately predict human behavior or intentions, particularly in cases involving malicious intent, potentially leading to misjudgments in risk assessment.

False positives and false negatives represent another limitation, as LLMs may produce erroneous risk assessments, either causing unnecessary concern or overlooking actual risks. Moreover, interpretability remains a challenge, as understanding the rationale behind LLM-generated risk assessments can be intricate, making it challenging to explain and justify the results to stakeholders. Over-reliance on LLMs for risk assessment, often driven by automation bias, may result in a lack of human oversight, potentially leading to the oversight of critical risks that necessitate human judgment. Lastly, LLM-based risk assessments may not fully align with evolving regulatory requirements and standards, which could expose organizations to legal risks.

The approach also cannot perceive whether the mitigations applied can help address the risk or limit the risk, as LLMs do not have essential mechanisms to evaluate them. These limitations underscore the need for a risk or domain expert augmented approach for risk assessment thereby enabling better and faster ways to identify risks and mitigate them before they end up causing harm. Further use of such python packages improves opportunity for the developers to identify mitigations early in their development process.

### 4.2 Way forward

To overcome some of the limitations with Large Language Model (LLM)-based risk assessment for downstream applications, a set of strategic approaches are proposed. These strategies aim to bolster the accuracy, dependability, and resilience of risk assessments. Firstly, fine-tuning techniques tailored to the specific domain of application can be implemented. This involves refining LLMs to better comprehend specialized knowledge and intricate contextual factors pertinent to the use case. Rigorous validation of these fine-tuned models against real-world scenarios is crucial to ascertain their effectiveness in risk assessment. Collaboration with domain experts and stakeholders should be an ongoing, iterative process to continually enhance model performance. Additionally, leveraging AI incident databases, AI vulnerability databases, and other relevant data sources can provide valuable insights to inform the assessment of risks in downstream applications.

Furthermore, augmenting LLM-based risk assessments with domain-specific expertise is a critical need. Engaging domain experts possessing specialized knowledge can help identify risks that LLMs may overlook due to their lack of domain-specific insight. These experts play a pivotal role in refining risk categorizations and aiding in the interpretation of LLM-generated assessments. Their contributions extend to providing information regarding additional risks to consider, defining the organization's risk tolerance, and offering domain-sensitive context, all of which enhance the meaningfulness of risk assessments.

Establishing a collaborative framework that combines the strengths of LLM capabilities with expert input ensures a more comprehensive and dependable risk assessment process. Moreover, enhancing the interpretability and explainability of LLM-generated risk assessments is paramount. The development of methods, including model interpretability tools and human-understandable explanations for risk assessments, facilitates transparent communication with stakeholders, promoting their comprehension and trust in the assessment outcomes.

## 5   CITING RELATED WORK

The paper has leveraged several works regarding risk and harms taxonomy including [50] , [49] , [12] , [25], [26] , [43], [38]  and [10]  to compile the list of broad risks and sub-risks.

### ACKNOWLEDGMENTS

This project is not supported by any grants.

## 6   HISTORY DATES

The submission is not currently presented to any journals.

https://doi.org/10.6028/NIST.SP.1270-draft